# High Speed WDM Interconnect Using Silicon Photonics Ring Modulators and Mode-Locked Laser

J. Müller[(1)], J. Hauck[(1)], A. Moscoso-Mártir[(1)], N. Chimot[(2)], S. Romero-García[(1)], B. Shen[(1)], F. Merget[(1)], F. Lelarge[(2)], J. Witzens[(1)]

[(1)] Institute for Integrated Photonics, Sommerfeldstr. 24, 52074 Aachen, Germany
([jmueller@iph.rwth-aachen.de](jmueller@iph.rwth-aachen.de))
[(2)] III-V Lab, Route de Nozay, 91461 Marcoussis, France

**Abstract** *We demonstrate an 8 by 14 Gbps compatible WDM link based on a single-section semiconductor mode-locked laser, silicon photonics resonant ring modulators and joint channel reamplification with a semiconductor optical amplifier operated in the linear regime. Individual channels reach a data rate of 25 Gbps with signal quality-factors above 7.*

## Introduction

In this paper we investigate the feasibility of an integrated Wavelength Domain Multiplexed (WDM) Silicon Photonics (SiP) Transmitter (Tx) using a single-section semiconductor Mode-Locked Laser (MLL) and Resonant Ring Modulators (RRM), followed by joint channel re-amplification with a Semiconductor Optical Amplifier (SOA) after incurring laser coupling and modulator insertion losses (IL). Eight consecutive optical comb lines of a MLL with a 102.6 GHz free spectral range (FSR) are modulated at 14 Gbps with a signal quality factor (Q-factor) above 7 and are consequently compatible with a bit error rate (BER) below $1 \times 10^{-12}$. An individual channel supports 25 Gbps with the aforementioned signal quality. Since proof of principle results described here rely on a partially integrated Tx combined with external fibre-coupled components, and thus a larger number of optical interfaces than required in a fully integrated solution, as well as sub-optimum grating couplers (GC), better performance is expected from the next chip iteration. Required MLL performance to support an N×25 Gbps link is derived in terms of relative intensity noise (RIN) and minimum power per line.

## Experimental Setup

The first element of the setup (Fig. 1) is a quantum-dash single-section MLL developed and fabricated by III-V Lab.

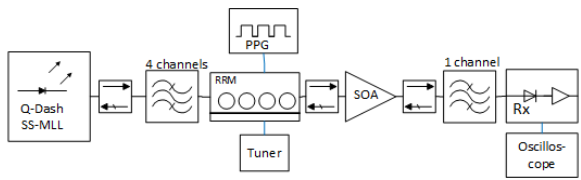

**Fig. 1:** Block diagram of the experimental setup

The buried ridge stripe Fabry-Perot laser has a 1.25 µm wide stripe and is based on a gain material consisting of six layers InAs Q-Dashes in an InGaAsP barrier grown on an InP wafer. The rear facet of the laser has a highly reflective coating, the front facet is as-cleaved. Based on the total laser RIN and the RF-linewidth, we selected a laser operating point[1] at 25°C and 238 mA injection current. The spectrum is centred at 1542 nm (Fig. 2) with a FSR of 102.6 GHz. 15 consecutive laser lines are within 3 dB of the maximum fibre-coupled line power of 1.5 dBm (measured after first isolator).

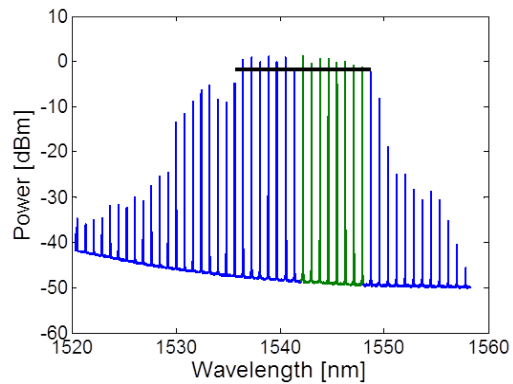

**Fig. 2:** Power spectrum of the laser (resolution bandwidth: 10 pm). The green lines were used for modulation

An isolator shields the laser from destabilizing back-reflections, and is followed by a 3 nm pass-band filter, which selects and transmits 4 lines downstream (IL=1 dB). Light is grating coupled to a SiP chip and modulated with a RRM using a phase shifter based on the plasma dispersion effect, with a 10 µm radius resulting in a 10 nm FSR and a loaded Q-factor of 5050. A PRBS-7 signal generated by a programmable pattern generator (PPG) with 2 $V_{pp}$ and -1 V reverse bias is applied to the RRM. Under these conditions, the maximum achievable Optical Modulation Amplitude (OMA) is estimated as -6.4 dB for our device[1] and the corresponding electro-optic -3 dBe cut-off frequency is 26 GHz.

We do however modify the detuning (difference between carrier frequency and RRM resonance frequency) of the RRM resonance to increase the extinction, at the cost of a reduced OMA and bandwidth (BW)[2], maximizing the signal Q-factor. Since noise in the link is dominated by RIN and amplified spontaneous emission (ASE) generated in the SOA, signal quality is dependent on extinction (typical detuning maximizing the signal Q-factor is -5.7 GHz for 14 Gbps resulting in BW=17.4 GHz, IL=7.2 dB, Ext.=12 dB, OMA=-7.5 dB, and is -8.3 GHz for 25 Gbps resulting in BW=18.8 GHz, IL=5.2 dB, Ext.=7 dB, OMA=-6.5 dB, where IL is defined as the waveguide to waveguide transmission loss in the '1'-bit state). The GCs at the input and output of the chip in combination with the waveguides leading to and from the RRM have a total loss of 10.3 dB. Light is reamplified with a commercial SOA (Thorlabs S9FC1004P). It has a small signal gain of 24 dB and a noise figure of 7.5 dBe measured at 1545 nm, including the effect of IL of isolators installed before and after the SOA. Finally, a single modulated line was filtered out with a single channel filter (BW=40 GHz, IL=2.4 dB) and sent to a 40 GHz photo-receiver (Rx, Finisar/U2T XPRV2021A). Eye diagrams were recorded on an Agilent DSA-X 920048 real time oscilloscope with the analog bandwidth throttled to 11 GHz for 14 Gbps measurements and set to 21 GHz for 25 Gbps measurements. While we are targeting an 8-channel system, only 4 channels could be jointly sent through the 3 nm tuneable filter used in the test bench, and two groups of adjacent channels had to be sequentially measured after changing the filter's centre frequency.

**Results and Analysis**

After measuring the aggregate laser RIN for isolated laser lines, integrated between 5 MHz and 20 GHz[1] (Fig. 3), we selected a group of eight adjacent lines with a comparatively high output power and low RIN and sequentially modulated them recording eye diagrams at 14 and 25 Gbps. All eight lines reached a signal Q-factor ($Q_{sig}$) above 7 at 14 Gbps, 1 line reached a Q-factor above 7 at 25 Gbps (Fig. 4).

When comparing Figs. 3 and 4, a correlation between $Q_{sig}$ and RIN and line power is readily apparent. Thus, we repeated the experiment with an instrument grade tuneable laser (negligible RIN) in order to characterize the influence of varying laser power on the system performance independently and calibrate a link performance model. Fig. 5 shows $Q_{sig}$ vs. laser power as recorded with the instrument grade laser and compares it to the model.

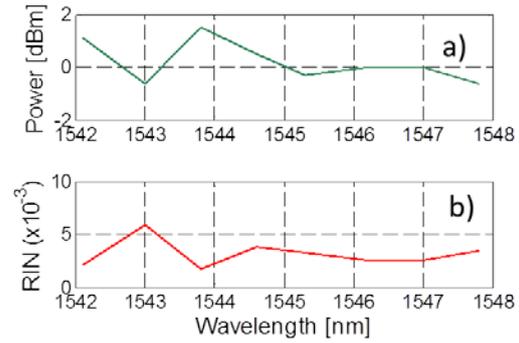

**Fig. 3:** a) Line Power and b) aggregate RIN per line as integrated from 5 MHz to 20 GHz

Interestingly, the optimum detuning with the MLL as light source was slightly smaller than for the instrument grade laser (detunings reported above), further increasing the extinction at the cost of some OMA reduction, presumably due to the additional sensitivity to extinction resulting from the laser RIN. At 14 Gbps, we maximized $Q_{sig}$ with the MLL for a detuning of -5 GHz, resulting in Ext.=15 dB, OMA=-7.9 dB, IL=7.8 dB. At 25 Gbps best results were obtained at -7 GHz detuning, resulting in Ext.=9 dB, OMA=-6.9 dB, IL=6.3 dB.

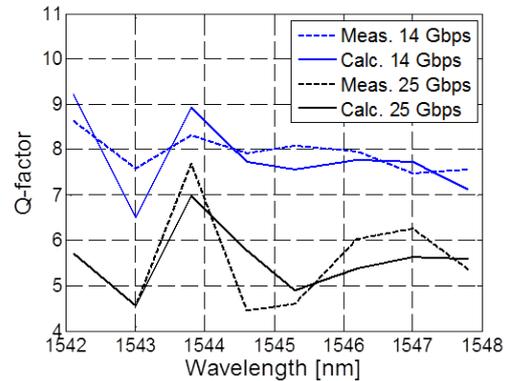

**Fig. 4:** $Q_{sig}$ at 14 (blue) and 25 Gbps (black) for each line

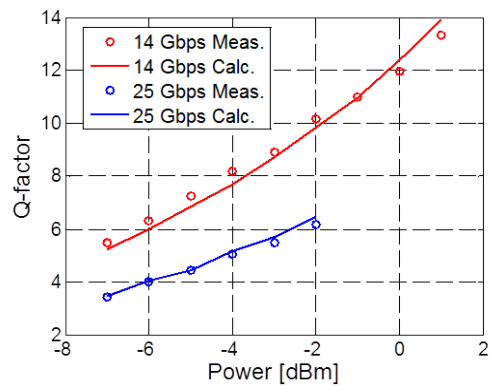

**Fig. 5:** Verification of analytical model by fitting the Q-factor as a function of varying instrument grade laser power

We modeled $Q_{sig}$ using following equation[1]

$$Q_{sig} = \eta(P_{1,Rx} - P_{0,Rx}) \times$$

$$\left(\sqrt{P_{n,Rx}^2 BW + 2G^2 F h f_0 P_{1,SOA} BW + RIN_t P_{1,Rx}^2} + \sqrt{P_{n,Rx}^2 BW + 2G^2 F h f_0 P_{0,SOA} BW + RIN_t P_{0,Rx}^2}\right)^{-1}$$

where $P_{1,Rx}$ and $P_{0,Rx}$ are the '1' and '0' bit power levels at the input of the Rx, $P_{n,Rx}$ is the input referred noise of the photo-receiver in W·Hz$^{-0.5}$ and $F$ is the noise factor of the SOA ($F=10^{NF/10}$). Furthermore, $h$ is Planck's constant, $f_0$ is the optical carrier frequency, $P_{1,SOA}$ and $P_{0,SOA}$ are the '1' and '0' power levels at the input of the SOA, $BW$ is the equivalent noise bandwidth of the Rx including the bandwidth setting of the oscilloscope. $RIN_t$ is the aggregate RIN (on a linear scale), $G$ is the total gain between the input of the SOA and the input of the photo-receiver and $\eta$ is the eye closure penalty due to inter-symbol interference (ISI) expected to occur due to the throttled oscilloscope bandwidth. The ISI penalty $\eta$ was extracted from the recorded eye diagrams[1] with values between -0.9 and -1 dB. Due to ISI, the '0' bit noise level was also higher than expected, which was taken into account in the calibrated model by increasing the value of $P_{0,SOA}$ in the equation. Finally, we had to assume $NF$=10.5 dBe in the model to account for higher measured noise levels. This discrepancy is still under investigation.

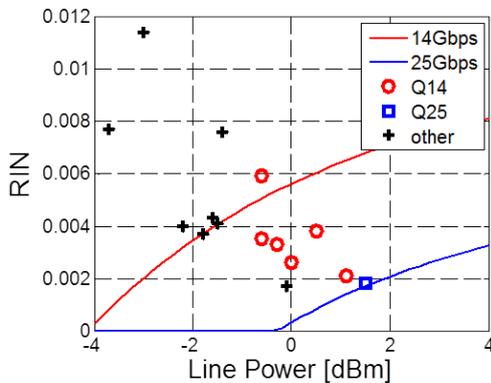

**Fig. 6:** Required and actual laser line performance

We applied the same model to the MLL system measurements including the independently measured RIN (Fig. 3(b)) and found it to predict the signal Q-factors very well. We also verified that in the reported data, Q-factors are limited by RIN and ASE, whereas Rx-noise plays a negligible role. Based on the calibrated link model, we derived the required laser line performance boundaries to maintain 10$^{-12}$ BER at 14 and 25 Gbps (Fig. 6) and compared them to the characteristics of the actual laser lines.

The circles represent lines for which 14 Gbps was achieved (Q>7), the square a line for which 25 Gbps was achieved (Q>7), and the crosses laser lines for which the Q-factor has not been measured. While it is apparent that most lines support 14 Gbps, significant improvement is still required to reliably reach 25 Gbps. Some improvement will arise from the SiP Tx since the 10.3 dB GC and waveguide losses can be straightforwardly decreased by at least 3 dB and hybrid laser integration may further create the opportunity of removing an additional optical interface. While we have already shown adequate RIN levels[1], we have yet to demonstrate them jointly with the increased power levels obtained here. The issue of laser and SOA isolation will also have to be addressed in a fully integrated solution.

A final sanity check consists in verifying the power budget after the SOA for the fully integrated solution. The utilized quantum well SOA has an output power of 15 dBm at the -3 dB gain compression point, corresponding to 6 dBm per channel for an 8-channel solution, which is in line with the required power roughly estimated in ref. 1. While the SOA needs to be generally operated in the linear regime when re-amplifying an already modulated carrier, encoding synchronized data channels to guarantee balancing of '1' and '0' levels across the channels would alleviate this situation and allow driving the SOA deeper into saturation. Additional encoding is however undesirable both from the point of view of excess power consumption and from the point of view of compatibility with existing standards (e.g., the requirement of synchronized channels), so that we are aiming at a system architecture supporting fully independent optical channels.

**Conclusions**

We have demonstrated an 8×14 Gbps compatible WDM link based on a single-section MLL, a SiP RRM and a SOA with signal Q-factors of 7 compatible with 10$^{-12}$ BER. A single channel reaches 25 Gbps with a Q-factor of 7. By modelling the link we have extrapolated the performance requirements to reliably support 25 Gbps signalling.

**Acknowledgements**

The authors acknowledge funding by the European Commission under contract numbers 619591, 293767 and 279770.

Chips were fabricated at Singapore's Institute of Microelectronics (IME).